\documentclass[preprint,aps,amsmath,amssymb]{revtex4}

\newcommand {\bc}{\begin{center}}
\newcommand {\ec}{\end{center}}
\newcommand {\bea}{\begin{eqnarray}}
\newcommand {\eea}{\end{eqnarray}}
\newcommand {\be}{\begin{equation}}
\newcommand {\ee}{\end{equation}}

\def\lsim{\mathrel{\rlap{\lower4pt\hbox{\hskip1pt$\sim$}}
    \raise1pt\hbox{$<$}}}               
\def\gsim{\mathrel{\rlap{\lower4pt\hbox{\hskip1pt$\sim$}}
    \raise1pt\hbox{$>$}}}

\usepackage[mathscr]{euscript}
\usepackage{graphicx}

\begin{document}


\title{Scale breaking and fluid dynamics in a dilute 
two-dimensional Fermi gas}

\author{Clifford Chafin and Thomas~Sch\"afer}

\affiliation{Department of Physics, North Carolina State University,
Raleigh, NC 27695}

\begin{abstract}
We study two observables related to the anomalous breaking of scale 
invariance in a dilute two dimensional Fermi gas, the frequency shift 
and damping rate of the monopole mode in a harmonic confinement potential. 
For this purpose we compute the speed of sound and the bulk viscosity of 
the two dimensional gas in the high temperature limit. We show that the 
anomaly in the speed of sound scales as $(2P-\rho c_s^2)/P\sim z/[\log(
T/E_B)]^2$, and that the bulk viscosity $\zeta$ scales as $\zeta/\eta 
\sim z^2/[\log(T/E_B)]^6$. Here, $P$ is the pressure, $c_s^2$ is the speed 
of sound, $\eta$ is the shear viscosity, $z$ is the fugacity, and $E_B$ is 
the two-body binding energy. We show that our results are consistent with 
the experimental results of Vogt et al.~[Phys.~Rev.~Lett.~108, 070404 
(2012)]. Vogt et al.~reported a frequency shift $\delta\omega/\omega$
of the order of a few percent, and a damping rate smaller than the 
background rate $\Gamma/\omega_0\sim 5\%$. 
\end{abstract}

\maketitle

\section{Introduction}
\label{sec_intro}
 
 Scale invariant or nearly scale invariant fluids play a role in 
many areas of physics. Examples include the three dimensional 
Fermi gas at unitary, the quark gluon plasma at very high temperature, 
and a number of fluids that can be described in terms of holographic 
dualities \cite{Schafer:2009dj,Adams:2012th}. A very interesting 
example is provided by a two-dimensional gas of fermions interacting 
via a zero range interaction. This system is scale invariant at 
the classical level, but scale invariance is broken in the quantum
theory. The quantum mechanical scattering amplitude depends logarithmically 
on a scale, which we can take to be the binding energy $E_B$ of the 
two-body bound state. This is analogous to what happens in QCD in 
three dimensions. QCD is classically scale invariant, but at the 
quantum level scale invariance is broken and the coupling depends
logarithmically on the QCD scale parameter. 

 Two properties related to the breaking of scale invariance were
recently studied by Vogt et al. \cite{Vogt:2011}. The first is the 
frequency of the monopole mode in a harmonically trapped gas. One can 
show that in a scale invariant gas this mode has frequency $2\omega_0$, 
where $\omega_0$ is the frequency of the harmonic confinement potential 
\cite{Pitaevskii:1997}. Deviations from this value provide a measure 
of scale breaking \cite{Taylor:2012,Hofmann:2012np}. Vogt et al.~found 
that these deviations are small, on the order of a few percent, for the 
entire range of parameters studied in their experiment. The second 
observable is the damping of the monopole mode. In a scale invariant 
fluid the monopole mode is undamped \cite{Castin:2011}. The experiments 
find the that the damping is too small to be reliably measured, although 
one has to keep in mind that the background damping rate is sizeable,
$\Gamma\simeq 0.05 \omega_0$.  

 In this work we present a rigorous calculation of the frequency
shift and the damping rate of the monopole mode in the high temperature
limit. Our calculation is based on the virial expansion for thermodynamic
properties, and on kinetic theory for non-equilibrium effects. We will
show that the results are in agreement with the measured frequency 
shift, and consistent with the failure of the experiment to observe
a non-zero damping rate. We note that the experimental data were taken
for $T/T_F\sim 0.4$ and a range of values of $\log(T_F/E_B)$. In the 
vicinity of the BCS/BEC crossover, corresponding to $\log(T_F/E_B)\sim 
0$, it is not clear that a high temperature calculation is quantitatively
reliable. Ultimately comparison between theory and experiment will
determine in what range of $T/T_F$ the kinetic theory description is 
applicable. In the case of the three-dimensional Fermi gas at unitarity 
there is some evidence that kinetic theory is reliable for $T\gsim 0.4 
T_F$, see for example \cite{Enss:2010qh}.

 Our study builds on earlier work that relates the frequency 
shift of the monopole mode to scale breaking in the speed of sound 
\cite{Taylor:2012}, and on our own work on bulk viscosity in the 
three dimensional Fermi gas \cite{Schaefer:2013oba}. In the latter
work we showed that the bulk viscosity of the three dimensional 
Fermi gas near unitarity scales as $\zeta\sim [(\Delta P)/P]^2 \eta$, 
where $\Delta P =P-\frac{2}{d}{\cal E}$ is the scale breaking part
of the pressure and $\eta$ is the shear viscosity. Here, $d$ is
the number of spatial dimensions and ${\cal E}$ is the energy
density. We will show that the bulk viscosity of the two dimensional
gas in the limit $T\gg E_B$ is even smaller than this estimate 
suggests. We find that $\zeta$ is suppressed by two additional 
powers of $\log(T/E_B)$. 

 This paper is organized as follows. In Sect.~\ref{sec_virial} introduce 
a diagrammatic approach to the virial expansion. We also apply this method 
to the calculation of the quasi-particle energy. In Sect.~\ref{sec_freq}
we compute the frequency of the monopole mode. In Sect.~\ref{sec_bulk}
we describe a calculation of the bulk viscosity in kinetic theory. We 
use the result to compute the damping of the monopole mode. We 
present an outlook in Sect.~\ref{sec_out}. 

\section{Equilibrium and quasi-particle properties}
\label{sec_virial}
\subsection{Two body interaction}
\label{sec_2bdy}

 A dilute gas of non-relativistic spin 1/2 fermions can be described 
by the effective lagrangian
\be 
\label{l_4f}
{\cal L} = \psi^\dagger \left( i\partial_0 +
 \frac{\nabla^2}{2m} \right) \psi 
 - \frac{C_0}{2} \left(\psi^\dagger \psi\right)^2 ,
\ee
where $m$ is the mass of the fermion and $C_0$ is the coupling 
constant. The scattering amplitude in the spin singlet channel is 
\be 
\label{Amp}
 {\cal A}(E)= \frac{1}{C_0^{-1}-\Pi(E)}\, ,
\ee
where $\Pi(E)$ is given by
\be 
\Pi(E)= \int \frac{d^2q}{(2\pi)^2} \frac{1}{E-\frac{q^2}{m}+i\epsilon}\, . 
\ee
This can be compared to the general structure of the s-wave
scattering amplitude in two dimensions \cite{Randeria:1990}
\be 
{\cal A}(E)= \frac{4\pi}{m}
  \frac{1}{-\pi\cot\delta(E)+i\pi}\, ,
\ee
where $\delta(E)$ is the s-wave scattering phase shift. We conclude 
that $\cot\delta(E)=\frac{1}{\pi}\log(\frac{E}{E_B})$ where $E_B$ is the 
two body binding energy. We also define the scattering length $a$ by 
$E_B=1/(ma^2)$. The relation between $C_0$ and $E_B$ depends on the
regularization scheme. In cutoff regularization we find
\be 
\label{C0_EB}
 \frac{1}{C_0(\Lambda)} = \frac{m}{4\pi}
  \log\left(\frac{mE_B}{\Lambda^2}\right)\, . 
\ee

\subsection{Thermodynamic potential}
\label{sec_omega2}
 
 In order to compute the thermodynamic potential it is useful to 
apply a Hubbard-Stratonovich transformation to the effective 
lagrangian. Introducing a complex di-fermion field $\Phi$ we 
can write
\be
\label{l_4f_hs}
{\cal L} = \psi^\dagger \left( i\partial_t +
 \frac{\nabla^2}{2m} \right) \psi  
 + \left[ (\psi\sigma_+\psi)\Phi + {\it h.c.}\right]
+\frac{1}{C_0}|\Phi|^2\, ,
\ee
where $\sigma_+$ is the Pauli spin raising matrix. The integration 
over the fermion fields is Gaussian. We obtain an effective 
action for the bosonic field $\Phi$,
\be
\label{s_ng_eff}
 S= -{\rm Tr}\left[\log\left(G^{-1}\left[\Phi,\Phi^*\right]\right)\right]+
     \int d^4x\, \frac{1}{C_0}|\Phi|^2 \, . 
\ee
where $G^{-1}$ is a $2\times 2$ matrix
\be
\label{ng_prop}
 G^{-1}\left[\Phi,\Phi^*\right] = 
 \left(\begin{array}{cc}
     i\partial_t + \frac{\nabla^2}{2m}  & \Phi^* \\
     \Phi & i\partial_t -\frac{\nabla^2}{2m}
 \end{array}\right).
\ee
The thermodynamic potential $\Omega$ is computed using the Matsubara 
formalism. We continue the fields to imaginary time $\tau$ and impose 
periodic/anti-periodic boundary conditions on the bosonic/fermionic 
fields. We also introduce a chemical potential for $\psi$. We evaluate 
the partition function by expanding the logarithm in powers of $\Phi$. 
The leading term is the free fermion loop, 
\be
\label{Om_1}
\Omega_1 = \frac{\nu z T}{\lambda^2} \left(
  1 - \frac{z}{2} + O(z^2) \right)\, ,
\ee
where $\nu=2$ is the number of degrees of freedom, $z=\exp(\mu/T)$ 
is the fugacity, and $\lambda=[(2\pi)/(mT)]^{1/2}$ is the thermal
wave length. Terms of order $z^2$ and higher arise from quantum statistics. 
The complete $O(z^2)$ result includes quadratic fluctuations in $\Phi$.
We find
\be 
\label{Omega_2}
\Omega_2 = T\sum_n\int \frac{d^2q}{(2\pi)^2}
   \log\left[ {\cal D}^{-1}(i\omega_n,q)\right]\, , 
\ee
where $\omega_n=2\pi n T$ are bosonic Matsubara frequencies
and ${\cal D}^{-1}(\omega_n,q)$ is the one loop di-fermion
polarization function
\be 
\label{chi_pp}
 {\cal D}^{-1}(i\omega_n,q) = \int \frac{d^2k}{(2\pi)^2} 
    \Bigg\{ \frac{1-f_k-f_{k+q}}{i\omega_n -\xi_k-\xi_{k+q}}
       - \frac{1}{E_B-2\epsilon_k} \Bigg\} \, .
\ee
Here, $f_k=[\exp(\beta\xi_k)+1]^{-1}$ is the Fermi-Dirac
distribution, $\xi_k=\epsilon_k-\mu$ and $\epsilon_k=k^2/(2m)$. 
In order to deriving equ.~(\ref{chi_pp}) we have used the relation 
between $C_0$ and $E_B$ given in equ.~(\ref{C0_EB}). The sum over 
Matsubara frequencies in equ.~(\ref{Omega_2}) can be performed
using contour integration. The result can be expressed in terms 
of the discontinuity of ${\cal D}^{-1}(\omega,q)$ along the real 
axis in the complex frequency plane. We obtain
\be 
\label{Omega_cut}
\Omega_2 =\frac{1}{2\pi i}\int_{-\infty}^{\infty} d\omega
  \int \frac{d^2k}{(2\pi)^2}\,{\it disc} 
  \left[\log {\cal D}^{-1}(\omega+i\epsilon,k)\right] \, f_{BE}(\omega)\, , 
\ee
where $f_{BE}(\omega)=[\exp(\beta\omega)-1]^{-1}$ is the Bose-Einstein 
distribution function. In order to compute $\Omega_2$ at second order in 
fugacity $z$ we need to evaluate ${\cal D}^{-1}(\omega,k)$ to zeroth order 
in $z$. We get 
\be 
\label{chi_0}
 {\cal D}^{-1}(\omega,k) = \frac{m}{4\pi} 
  \log\left(- \frac{\omega-\frac{\epsilon_k}{2}+2\mu}{E_B}\right)\, . 
\ee
The simplest strategy to compute the integral over $\omega$ and $k$ 
is to compute $n=(\partial\Omega)/(\partial\mu)$, and then integrate 
over $\mu$. The result can be used to extract the interaction part of 
the second virial coefficient. We find
\be 
\label{b_2}
\delta b_2 = e^{\beta E_B} 
   - 2 \int \frac{dk}{k}
   \frac{e^{-2\beta\epsilon_k}}
     {\left[\log(a^2k^2)\right]^2+\pi^2} \, , 
\ee
with $\beta=1/T$. The result agrees with the expectation from the 
standard Beth-Uhlenbeck expression for the second virial coefficient
in terms of the phase shift, 
\be 
 \delta b_2 = e^{\beta E_B}  + 
  \frac{1}{\pi}\int dk\, \left(\frac{d\delta}{dk}\right)
    \, e^{-2\beta\epsilon_k}\, .
\ee
The integral can be computed in terms of a function called 
$\nu(x)$ in \cite{Gradshteyn}. We find 
\be
\label{b2_nu}
\delta b_2(T) = \nu \left(\beta E_B\right)\, , 
 \hspace{1cm}
\nu(x) = \int_0^\infty \frac{x^t\,dt}{\Gamma(t+1)}\, . 
\ee
For small $x$ the function $\nu(x)$ can be expanded in 
inverse powers of $\log(1/x)$, see App.~\ref{app_nu} 
and \cite{Erdelyi}. This expansion determines the virial 
coefficient in the limit $T\gg E_B$. We get 
\be 
\delta b_2(T) = \frac{1}{\log(T/E_B)}
  +  \frac{\gamma_E}{[\log(T/E_B)]^2} + \ldots \, , 
\ee
where $\gamma_E$ is Euler's constant.

\subsection{Quasi-particle properties}
\label{sec_qp}

 We can construct a systematic expansion for the two-dimensional
gas in the dilute limit by writing the lagrangian in terms of 
fermion and boson degrees of freedom. For this purpose we 
write the lagrangian as the sum of free and interacting terms, 
${\cal L}= {\cal L}_0 + {\cal L}_1$, with
\bea
\label{L_01}
{\cal L}_0  &=&  \psi^\dagger {\cal S}^{-1}(\omega,p) \psi  
 + \Phi^*{\cal D}^{-1}(\omega,p)\Phi \, ,  \\
{\cal L}_1  &=& \left[ (\psi\sigma_+\psi)\Phi + {\it h.c.}\right]
 + \Phi^*\left[ C_0^{-1} - {\cal D}^{-1}(\omega,p)\right]\Phi
   \, . \nonumber 
\eea
Here, ${\cal S}(\omega,p)=[\omega-\xi_p]^{-1}$ is the fermion 
propagator and ${\cal D}(\omega,p)$ is the boson propagator given 
in equ.~(\ref{chi_0}). The bosonic term in the interaction serves
as a counterterm that removes fermion loop insertions in the 
boson self energy order by order in the expansion. The leading
order fermion self energy is given by 
\be 
 \Sigma(i\omega_m,k) = T\sum_n \int \frac{d^2q}{(2\pi)^2}
    {\cal D}(i\omega_n+i\omega_m,q+k)
    {\cal S}(i\omega_n,q)\, . 
\ee
The Matsubara sum can be performed as
in the previous section. The resulting contour integral receives
contributions from the cut in the boson propagator and the pole 
in the fermion propagator. At leading order in the fugacity expansion
we can neglect the cut contribution. In order to compute the 
quasi-particle self energy we analytically continue the pole term
to the on-shell point $i\omega_m=\xi_k$. The result can be 
determined analytically as an expansion in $[\log(T/E_B)]^{-1}$. 
We get 
\be 
\label{sigma_k}
\Sigma(k) = -\frac{2zT}{\log(T/E_B)} \left\{ 1 -
 \frac{1}{\log(T/E_B)} \left[ 
   - {\it Ei}\left(-\frac{\epsilon_k}{T}\right)
   + \log\left(-\frac{\epsilon_k}{2T}\right)- i\pi \right]
   + \ldots \right\}\, ,
\ee
where ${\it Ei}$ is the exponential integral. The quasi-particle
energy is $E_k=\epsilon_k+{\it Re}\Sigma(k)$ and the width is 
$\Gamma_K=-{\it Im}\Sigma(k)$. We note that ${\it Re}\Sigma$ is 
momentum independent at leading order in $[\log(T/E_B)]^{-1}$,
but develops momentum dependence at next-to-leading order.

\section{Frequency shift of the monopole mode}
\label{sec_freq}

 We follow the work of \cite{Taylor:2008,Taylor:2009,Taylor:2012}
and compute the frequency of the monopole mode using a variational
method. The use of variational methods in hydrodynamics was pioneered
in \cite{Zilsel:1950,Ito:1953}. The Euler equations follow from the 
variational principle $\delta S = 0$ with $S=\int dt\, L$ and 
\be
 L = \int d^2r \, \bigg[ \frac{1}{2}\rho\vec{u}^{\,2}
     - {\cal E}(\rho,\bar{s})
     - \frac{\rho}{m} V_{\it ext}(r) 
   -\phi \bigg( \frac{\partial \rho}{\partial t} 
           + \vec\nabla\cdot\left(\vec{u}\, \rho\right)\bigg)
   -\xi  \bigg( \frac{\partial\rho\bar{s}}{\partial t} 
           + \vec\nabla\cdot\left(\vec{u}\rho\,\bar{s}\right)\bigg)
   \bigg]\, . 
\ee 
The hydrodynamic variables are the mass density $\rho$, the velocity 
$\vec{u}$, and the entropy per particle $\bar{s}=s/n$. Here, $s$ is 
the entropy density and $n$ is the particle density. The fields 
$\phi$ and $\xi$ are Lagrange multipliers that enforce the continuity 
and entropy conservation equations. ${\cal E}(\rho,\bar{s})$ is 
the energy density and $V_{\it ext}$ is an external potential. We 
will consider a harmonic, rotationally invariant potential 
$V_{\it ext}=\frac{1}{2}m\omega_0\vec{r}^2$. We also add Lagrange 
multipliers $\mu_0$ and $T_0$ to allow for solutions with finite 
particle number and entropy, 
\be 
L' = L + \frac{1}{m}\int d^2r\, \left( \rho\mu_0 
  + \rho\,\bar{s}\,T_0 \right)\, . 
\ee
Varying $S'$ with respect to the hydrodynamic variables and
setting $\vec{u}=0$ we get the hydrostatic equations
\bea 
\label{hstat1}
\vec{\nabla} P &=&
  \left.\frac{\partial P}{\partial\rho}\right|_{\bar{s}}
     \vec{\nabla}\rho  
 +\left.\frac{\partial P}{\partial\bar{s}}\right|_{\rho}
     \vec{\nabla}\bar{s} 
 = -n \vec{\nabla} V_{\it ext}\, ,  \\
\label{hstat2}
\vec{\nabla} T &=&
  \left.\frac{\partial T}{\partial\rho}\right|_{\bar{s}}
     \vec{\nabla}\rho  
 +\left.\frac{\partial T}{\partial\bar{s}}\right|_{\rho}
     \vec{\nabla}\bar{s} 
 = 0\, . 
\eea
The partial derivatives are related by the Maxwell relation
\be 
\label{Maxwell}
 \rho\left.\frac{\partial T}{\partial\rho}\right|_{\bar{s}}
 = \frac{1}{\rho}\left.\frac{\partial P}{\partial\bar{s}}\right|_{\rho}
 \, . 
\ee
We will denote the solution to the hydrostatic equations by $\rho_0$ 
and $\bar{s}_0$.  Small oscillations around this solution are governed by 
\be 
 L_2 = \frac{1}{2}\int d^2r\, \bigg[ \rho_0 \vec{u}^{\,2}
  -\frac{1}{\rho_0}\left(\frac{\partial P}{\partial\rho}\right)_{\!\bar{s}}
     (\delta\rho)^2
  -2\rho_0 \left(\frac{\partial T}{\partial\rho}\right)_{\!\bar{s}}
     \delta\rho \delta\bar{s}
  - \rho_0 \left(\frac{\partial T}{\partial\bar{s}}\right)_{\!\rho}
     (\delta\bar{s})^2
\bigg] \, .
\ee
A variational ansatz for monopole vibrations is given by $\vec{u}=
\vec{u}_0 e^{-i\omega t}$ with $\vec{u}_0=\alpha \vec{r}$. This 
ansatz corresponds to an exact solution of the Euler equation in 
the scale invariant case. The continuity equation implies  
that 
\be 
\delta\rho = -\frac{i}{\omega}\vec{\nabla}\cdot\left(\rho_0\vec{u}_0\right)
\, ,\hspace{0.5cm}
\delta\bar{s} = -\frac{i}{\omega}\vec{u}_0\cdot \vec{\nabla} 
   \left(\bar{s}_0\right) \, .
\ee
The variational estimate for the mode frequency $\omega$ is obtained
by setting $L_2$ to zero. Using the hydrostatic equations 
(\ref{hstat1}-\ref{hstat2}) and the Maxwell relation (\ref{Maxwell})
we find \cite{Taylor:2012}
\be 
 \frac{\omega^2}{4\omega_0^2} = 
 1 - \frac{1}{2} \bigg[\int d^2r\, \gamma_2(r)\bigg] \,\bigg/\,
                 \bigg[\int d^2r\, n_0(r)V_{\it ext}(r)\bigg]
\ee
with $\gamma_2=2P-\rho c_s^2$, where $c_s^2$ is the speed of 
sound. This result has a simple physical interpretation: The 
deviation of the monopole frequency from the value $\omega=2
\omega_0$ in a scale invariant theory \cite{Pitaevskii:1997}
is governed by the trap average of the difference between 
the actual speed of sound and the sound speed $c_{s,0}^2=2P/
\rho$ in a scale invariant gas.

\begin{figure}[t]
\bc\includegraphics[width=9cm]{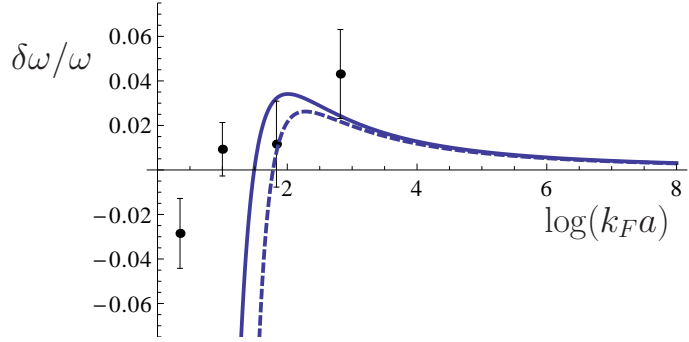}\ec
\caption{\label{fig_shift}
Frequency shift relative to the prediction $\omega=2\omega_0$ for 
a scale invariant fluid of the collective monopole mode in a two
dimensional Fermi gas. The frequency shift is shown as a function
of $\log(k_Fa)$ for $T/T_F=0.42$. The solid line is based on the virial
expansion, the dashed line shows the result at leading order in 
$1/[\log(T/E_B)]$, and the data are taken from Fig.~1c in \cite{Vogt:2011}.}   
\end{figure}

 In the dilute limit we can compute $\gamma_2$ using the 
virial equation of state. We find
\be 
\gamma_2 = \frac{mz^2E_B^2}{\pi}\, \nu^{\prime\prime}
   \left(\frac{E_B}{T}\right) \, . 
\ee
We perform the trap average using the density profile in 
the high temperature limit, 
\be 
n_0(x) = \frac{mT}{2\pi} \left( \frac{T_F}{T}\right)^2 
  \exp\left(-\frac{m\omega_0^2r^2}{2T}\right)\, , 
\ee 
where $T_F=\sqrt{N}\omega_0$ is the Fermi temperature of the 
trap. We can now compute the frequency shift in the dilute 
limit
\be
 \frac{\omega^2}{4\omega_0^2} = 
 1 - \frac{1}{8} \frac{T_F^2E_B^2}{T^4} \, 
   \nu^{\prime\prime}\left(\frac{E_B}{T}\right)\, . 
\ee
The result is shown in Fig.~\ref{fig_shift}. We observe that the 
frequency shift is in agreement with the data for $\log(k_Fa)\gsim 
1.75$. For smaller values of $\log(k_Fa)$ the contribution of the 
bound state is large and the virial expansion breaks down. In 
Fig.~\ref{fig_shift} we also show the contribution of the leading
$[\log(T/E_B)]^{-1}$ term in the second virial coefficient. We
observe that the leading term dominates for $\log(k_Fa)\gsim 2$.

\section{Bulk viscosity and the damping of the monopole mode}
\label{sec_bulk}
\subsection{Chapman-Enskog expansion}

 In the context of hydrodynamics the damping of the monopole 
mode is determined by the bulk viscosity of the two dimensional 
Fermi gas. In this section we will use kinetic theory and the 
Chapman-Enskog expansion to compute the bulk viscosity. An 
analogous calculation of the shear viscosity and the damping 
of the quadrupole mode can be found in 
\cite{Schafer:2011my,Bruun:2011,Baur:2013}.

 In kinetic theory we write the stress tensor of the gas in 
terms of the quasi-particle distribution function $f_p(\vec{x},t)$.
We will assume that the gas is spin-symmetric so that $f_p^\uparrow
=f_p^\downarrow\equiv f_p$. Close to equilibrium we can write 
\be 
 f_p(\vec{x},t) = f_p^0(\vec{x},t) + \delta f_p(\vec{x},t)
  = f_p^0(\vec{x},t) \left( 1 - \frac{\psi_p}{T} \right)\, , 
\ee
where $f_p^0(\vec{x},t)$ is the local equilibrium distribution in 
a fluid with local velocity $\vec{u}(\vec{x},t)$, temperature 
$T(\vec{x},t)$ and chemical potential $\mu(\vec{x},t)$. In the 
case of a fluid undergoing a scaling expansion the off-equilibrium 
factor has the form $ \psi_p= \chi^B(\vec{p}) \vec{\nabla}\cdot 
\vec{u}$. The off-equilibrium distribution $\delta f_p$ induced
by the bulk stress $\vec{\nabla}\cdot \vec{u}$ is determined by 
the Boltzmann equation 
\be 
\label{Beq}
  Df_p \equiv 
 \left( \frac{\partial}{\partial t} 
  + \vec{v}_p \cdot \vec{\nabla}_x
  + \vec{F} \cdot \vec{\nabla}_p \right)
f_p\left(\vec{x},t\right)
 = C[f_p]\, .
\ee
Here, $\vec{v}_p=\vec{\nabla}_pE_p$ is the quasi-particle velocity, 
$\vec{F}=-\vec{\nabla}_xE_p$ is the force term, and $C[f_p]$ is the 
collision term. Using the methods described in \cite{Schaefer:2013oba}
the left hand side of the Boltzmann equation can be written as 
\be
\label{stream}
\frac{T}{f_0} D f_0 =  \Bigg\{
  \frac{\alpha \rho c_T^2}{c_V} h - mc_s^2
  + \left[ \frac{1}{2}\vec{p}\cdot\vec{\nabla}_p 
             - \frac{\alpha \rho c_T^2}{c_V} 
  + \rho c_s^2 \left.\frac{\partial}{\partial P}\right|_T
             + \frac{\alpha \rho c_T^2}{c_V}T  
                      \left.\frac{\partial}{\partial T}\right|_P
\right] E_p\Bigg\}\, \vec{\nabla}\cdot \vec{u} \, . 
\ee
where $h$ is the enthalpy per particle, $c_s$/$c_T$ are the speed 
of sound at constant entropy per particle/temperature, $c_V$ is 
the specific heat at constant volume, and $\alpha$ is the thermal 
expansion coefficient. This result can be simplified by
writing $E_p=\epsilon_p+\Delta E_P$ and dropping terms of 
order $z^2$. We get
\bea
\frac{T}{f_0} D f_0 &=& \mbox{} \Bigg\{
  \frac{\alpha \rho c_T^2}{c_V} h - mc_s^2
  + \left[ 1 
          - \frac{\alpha \rho c_T^2}{c_V} \right]\epsilon_p \nonumber \\
\label{stream2}
  & & \label{lhs_2} \mbox{} + 
   \left[\frac{1}{2}\vec{p}\cdot\vec{\nabla}_p 
    +  \mu\frac{\partial}{\partial \mu}
    +  T\frac{\partial}{\partial T} -  1 \right] \Delta E_p
      \Bigg\}\, \vec{\nabla}\cdot \vec{u} \, . 
\eea
This result satisfies a number of consistency checks. Equ.~(\ref{stream2})
vanishes for a free gas, and for a general scale invariant gas 
characterized by a temperature independent second virial coefficient
and a scale invariant dispersion law of the form $\Delta E_p \sim 
zTg(\epsilon_p/T)$ where $g(x)$ is an arbitrary function. 

 In order to solve the Boltzmann equation in the limit $T\gg E_B$ 
we use the second virial coefficient given in equ.~(\ref{b2_nu}) 
and the quasi-particle energy using equ.~(\ref{sigma_k}). We expand 
all quantities to leading non-trivial order in $[\log(T/E_B)]^{-1}$. 
We get $\frac{T}{f_0}D f_0 \equiv X_p (\vec{\nabla}\cdot\vec{u})$ with
\be
\label{X_p}
 X_p =  \frac{2zT}{\left[\log(T/E_B)\right]^3}
  \left\{ \frac{\epsilon_p}{T} - \left( 1 + 2\gamma_E\right) 
     + 2\left(  {\it Ei}\left(- \frac{\epsilon_p}{T} \right)
                   -\log\left( \frac{\epsilon_p}{2T} \right)\right) 
 \right\}\, ,
\ee
where ${\it Ei}(x)$ is the exponential Integral. We note that 
both the thermodynamic terms as well as the self energy 
contain contributions of order $[\log(T/E_B)]^{-2}$, but these
terms cancel and $X_p$ scales as $[\log(T/E_B)]^{-3}$. Equ.~(\ref{X_p})
satisfies two sum rules
\be 
\label{ortho_3}
 \int d\Gamma_p \, f^0_p X_p = 0\, , \hspace{0.5cm}
 \int d\Gamma_p \, f^0_p \epsilon_p X_p = 0 \, , 
\ee
with $d\Gamma_p=d^2p/(2\pi)^2$. The sum rules follow from 
conservation of particle number and energy.

\subsection{Collision term}

The linearized collision operator can be written as
\be
 C[f^0_p+\delta f_p]\equiv \frac{f^0_p}{T} C_L[\chi_B(p)]
  \left(\vec{\nabla}\cdot\vec{u}\right)
\ee
where, at leading order in the fugacity, the collision term is 
dominated by two-body collisions. We have
\be 
 C_L[\chi_B(p_1)] = \int \Big(\prod_{i=2}^4d\Gamma_{i}\Big)
   w(1,2;3,4) f^0_{p_2}
   \left[\chi_B(p_1)+\chi_B(p_2)-\chi_B(p_3)+\chi_B(p_4)\right]\, ,
\ee
where $w(1,2;3,4)$ is the transition rate 
\be 
w(1,2;3,4) = (2\pi)^3\delta^2\Big(\sum_i \vec{p}_i\Big)
  \delta \Big( \sum_i E_{i}\Big) \left| {\cal A}\right|^2\, ,
\ee
and ${\cal A}$ is the scattering amplitude given in equ.~(\ref{Amp}). 
In order to be consistent with the calculation of the streaming
term we expand the scattering amplitude to leading order in 
$[\log(T/E_B)]^{-1}$. We get 
\be
 {\cal A} = \frac{4\pi}{m\log(T/E_B)}\, . 
\ee
To leading order in $z$ we can approximate the quasi-particle
energy by the non-interacting result $E_p\simeq \epsilon_p$. The 
linearized Boltzmann equation 
\be 
\label{BE_lin}
  X_p = C_L[\chi_B(p)]
\ee
can be solved by expanding the off-equilibrium factor $\chi_B(p)$ 
in Laguerre polynomials
\be 
\label{chi_exp}
 \chi_B(p) = \sum_{i=2}^N c_i L^{0}_i
   \left(\frac{\epsilon_p}{T} \right) \, .
\ee
Restricting the sum to terms of order $i\geq 2$ guarantees that 
the orthogonality constraints (\ref{ortho_3})  are satisfied. The 
coefficients $c_i$ can be determined by taking moments of the Boltzmann 
equation (\ref{BE_lin}) with  $L^0_i(\epsilon_p/T)$ for $i=2,\ldots,N$. 
As a first approximation we can take $N=2$. We find
\be
\label{chi_sol}
 \chi_B(p) = \frac{1}{4\pi\log(T/E_B)}
   \left[ 2 - 4\left(\frac{\epsilon_P}{T}\right) 
     + \left(\frac{\epsilon_P}{T}\right)^2  \right] \, . 
\ee
The convergence of the expansion in Laguerre polynomials was
studied in the case of three dimensions in \cite{Schaefer:2013oba}.
We found that corrections to the leading term contribute to 
the bulk viscosity at a level of less than 5\%. 

\subsection{Off-equilibrium stress and bulk viscosity}

 In fluid dynamics the trace of the stress tensor in the fluid rest 
frame is given by $\Pi\equiv\frac{1}{2}\Pi_{ii}=P-\zeta(\vec{\nabla}
\cdot\vec{u})$, where $\zeta$ is the bulk viscosity. In kinetic theory
this expression has to be matched against  
\be 
\label{bulk_kin}
\Pi[f_p] = \frac{\nu}{2}
\int d\Gamma_p\, \left( \vec{v}_p\cdot\vec{p}\cdot\right) f_p
 +  \nu\int d\Gamma_p \, E_p f_p - {\cal E}[f_p]\, , 
\ee
where $\nu=2$ is the number of spin degrees of freedom, $d\Gamma_p=
d^2p/(2\pi)^2$ is the phase space measure, $f_p=f(\vec{p},\vec{x},t)$ 
is the quasi-particle distribution function, $E_p$ is the quasi-particle 
energy and $\vec{v}_p=\vec{\nabla} E_p$ is the quasi-particle velocity. 
We split the distribution in an equilibrium and a non-equilibrium piece, 
$f_p=f_p^0+\delta f_p$, and write the bulk stress as
\be 
\label{bulk_kin_2}
 \Pi[f_p^0+\delta f_p] \equiv  \Pi[f_p^0]+\delta\Pi
  \equiv \Pi^0 + \delta\Pi\, . 
\ee
The term $\delta\Pi$ is then identified with viscous correction
$-\zeta(\vec{\nabla}\cdot\vec{u})$ in fluid dynamics. We compute 
$\Pi[f_p^0+\delta f_p]$ by functionally expanding equ.~(\ref{bulk_kin})
in powers of $\delta f_p$. We find \cite{Schaefer:2013oba}
\be 
\label{del_pi_fin}
\delta\Pi = \nu \int d\Gamma_p \, \delta f_p
  \left(\frac{1}{2}\vec{p}\cdot\vec{\nabla}_p 
    + \mu\frac{\partial}{\partial \mu}
    + T\frac{\partial}{\partial T} - 1 \right) \Delta E_p\, . 
\ee
It is interesting to note that the bulk stress is determined by 
the same scale violating part of the self energy that appears
in the streaming term (\ref{stream2}). This ensures that in 
a scale invariant gas there is no bulk viscosity irrespective 
of the structure of the off-equilibrium distribution function. 

 We can now compute the bulk viscosity by inserting the solution
of the linearized Boltzmann equation given in equ.~(\ref{chi_sol}) 
into the expression for the bulk stress. Comparing the result to 
$\delta\Pi=-\zeta(\vec{\nabla}\cdot\vec{u})$ determines the bulk
viscosity. We find
\be 
\label{zeta}
\zeta = \frac{1}{2\pi} \frac{z^2\lambda^{-2}}{[\log(T/E_B)]^4}\, .
\ee
This result is valid in the limit $z\ll 1$ and $\log(T/E_B)\gg 1$. 
Higher order corrections in $z$ require a calculation of the 
pressure at the level of the third virial coefficient, and the 
inclusion of three-body scattering in the collision term. Higher 
order terms in $[\log(T/E_B)]^{-1}$ can be determined by computing
the self energy to all orders in in the logarithm of $T/E_B$, as 
we have done for the second virial coefficient. However, unless we 
include bosonic quasi-particles in the kinetic theory, the result for
the bulk viscosity will still break down for $T\sim E_B$. We can 
compare equ.~(\ref{zeta}) to the result for the shear viscosity 
obtained in \cite{Schafer:2011my,Bruun:2011}
\be 
\eta = \frac{\lambda^{-2}}{\pi} \,\left[\log(T/E_B)\right]^2\,  , 
\ee
where we have taken the limit $\log(T/E_B)\gg 1$ used in the 
calculation of the bulk viscosity. We observe that $\zeta$ is 
suppressed by two additional powers of $[\log(T/E_B)]^{-1}$ 
compared to the expectation \cite{Schaefer:2013oba} $\zeta\sim 
[(\Delta P)/P]^2 \eta$, where $\Delta P=P-{\cal E}$ is the scale 
breaking part of the equilibrium pressure. The reason for this 
extra suppression is related to the fact that the leading scale 
violating term in the quasi-particle energy is just a shift in 
the chemical potential, which does not contribute to the bulk 
pressure. Finally, we can write equ.~(\ref{zeta}) in terms of 
dimensionless ratios. We find
\be
\frac{\zeta}{n} = \frac{1}{4\pi} \frac{1}{[\log(T/E_B)]^4}
   \left( \frac{T_F^{\it loc}}{T}\right) \, , 
\ee
where $T_F^{\it loc}$ is the Fermi temperature of the homogeneous gas. 

\begin{figure}[t]
\bc\includegraphics[width=9cm]{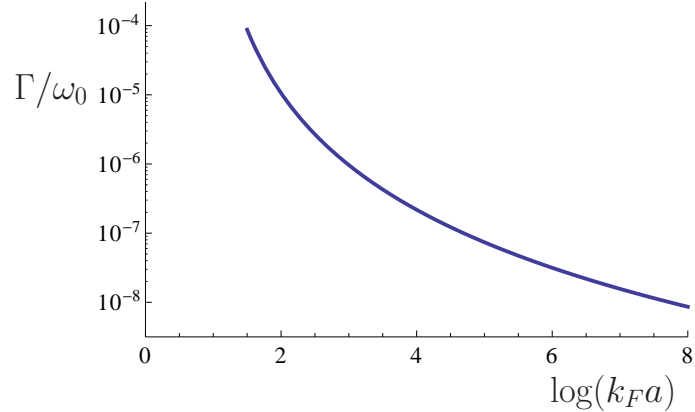}\ec
\caption{\label{fig_damp}
Damping of the collective monopole mode in a 2d Fermi gas. The
figure shows $\Gamma/\omega_0$, the damping rate in units of the 
trap frequency,  as a function of $\log(k_Fa)$ for 
$T/T_F=0.42$ and $N=4\cdot 10^3$.}   
\end{figure}

\subsection{Damping of monopole oscillations}

 The damping of the monopole mode is determined by the amount 
of energy dissipated by viscous effects. Note that the cloud remains
approximately isothermal so that thermal conductivity does not 
contribute to dissipation. The rate of energy dissipation is 
\be 
\dot{E} = -\frac{1}{2}\int d^2r\, \zeta(r) 
  \left(\vec{\nabla}\cdot\vec{u}\right)^2\, , 
\ee
and the damping constant is $\Gamma=-\dot{E}/(2E)$, where $E$ is 
energy of the collective mode. Using the velocity profile $\vec{u}
\sim \vec{r}$ of the monopole mode we find
\be 
\Gamma = \bigg[\int d^2r\, \zeta(r)\bigg] \,\bigg/\,
                 \bigg[m\int d^2x\, n_0(r)\vec{r}^{\, 2}\bigg]\, . 
\ee
This result can be evaluated in the same way as the frequency 
shift considered in Sect.~\ref{sec_freq}. We get 
\be 
\label{Gam_trap}
\frac{\Gamma}{\omega_0} = \frac{1}{32\pi N^{1/2}}
    \frac{1}{[\log(T/E_B)]^4}
    \left( \frac{T_F}{T}\right)
\ee
where $N$ is the number of particles and $T_F=\sqrt{N}\omega_0$ is 
the Fermi temperature of the trap. The result is plotted in 
Fig.~\ref{fig_damp}. We consider the conditions explored in the 
experiment of Vogt et al., $N=4\cdot 10^3$ and $T/T_F=0.42$. 
Based on our results for the frequency shift we assume that 
equ.~(\ref{Gam_trap}) is reliable for $\log(k_Fa)\gsim 2$. 
We observe that the damping constant in this regime is extremely 
small, $\Gamma/\omega_0 < 10^{-4}$. This is consistent with the 
measurements of Vogt et al., who find $\Gamma/\omega_0 < 5\cdot 10^{-2}$, 
but our result implies that it will be very difficult to measure the bulk
viscosity of the two dimensional gas in the BCS limit. 

\section{Outlook}
\label{sec_out}

 The main results obtained in this work are the scaling of the 
pressure anomaly $(\Delta P)/P\sim z/[\log(T/E_B)]^2$ and the 
bulk viscosity $\zeta/\eta \sim z^2/[\log(T/E_B)]^6$ in the 
high temperature limit $z\ll 1$ and $\log(T/E_B)\gg 1$. These
results quantitatively explain the observed frequency shift of 
the monopole mode, and the failure of the experiment to detect
a non-zero damping rate. 

 It would be interesting to extend the calculation to the regime
$T\sim E_B$. In the context of kinetic theory this would presumably
require the inclusion of explicit bosonic degrees of freedom. 
Alternatively, one might try to compute the pressure and the 
bulk viscosity using purely diagrammatic methods. At weak coupling 
the mechanism for generating bulk viscosity is related to a 
finite relaxation time for rearranging the internal energy of 
the gas among non-interacting and interacting terms in the single 
particle energy. For $T\sim E_B$ the physical mechanism is likely
to be related to the formation of molecules. This involves 
three-body collisions, and is therefore suppressed in the low 
density limit $z\ll 1$, but the process may be enhanced by powers 
of $\log(T/E_B)$. 

 It is also important to consider the frequency dependence of 
the bulk viscosity. Taylor and Randeria proved the sum rule 
\cite{Taylor:2012}
\be 
\frac{2}{\pi}\int_0^\infty d\omega\, \zeta(\omega) 
 = 3P-{\cal E}-\rho c_s^2\, . 
\ee
At high temperature the right hand side scales as $z^2mT^2/[\log(T
/E_B)]^3$. On the left hand side the contribution of the transport
peak is $(\Delta\omega)z^2mT/[\log(T/E_B)]^4$, where $\Delta\omega$ 
is the width of the transport peak. Consistency with the sum rule
then requires that $\Delta\omega\leq T \log(T/E_B)$. This bound
can be compared to the width of the transport peak in the shear 
channel, which is $\Delta\omega\sim zT/[\log(T/E_B)]^2\ll T$. 
Taylor and Randeria also studied the tail of the spectral function 
(see also \cite{Hofmann:2011qs}). They find $\zeta(\omega) \sim 
z^2\lambda^{-2}T/(\omega[\log(\omega/E_B)]^2[\log(T/E_B)]^2)$, 
where we have used the high temperature limit of the contact.
This result matches the kinetic theory prediction for $\omega\sim T$.

 Acknowledgments: We thank John Thomas for useful discussions, and 
E.~Taylor and M.~Randeria for communications regarding their work.
This work was supported in parts by the US Department of Energy 
grant DE-FG02-03ER41260.

\begin{appendix}
\section{Thermodynamics}
\label{app_th}

 In this appendix we compute thermodynamic properties like the scale 
breaking contributions to the equation of state and the speed of sound 
using the virial equation of state. We follow the methods used in 
\cite{Schaefer:2013oba}. At second order in the virial expansion we 
have $P=\frac{\nu T}{\lambda^{2}}(z+b_{2}(T)z^2)$. At this order 
the scale breaking contribution to the equation of state is 
\be
\frac{P-{\cal E}}{P} = - z Tb_2'(T)\, .
\ee
The enthalpy per particle is 
\be 
h= 2T \Big[1-z\Big(b_2(T)-\frac{1}{2}Tb_2^\prime(T)\Big)\Big]\, 
\ee
and the specific heats are given by
\bea
  c_{V} &=& \frac{\nu z}{\lambda^{2}}\left[1
      + z\Big( 2b_2(T)+T^{2} b^{\prime\prime}_{2}(T)\Big)\right]\, ,  
      \\ 
  c_{P} &=& c_{V}+\frac{\nu z}{\lambda^{2}}\Big[1+ 
      z\Big(4b_{2}(T) - 2T b^\prime_{2}(T)\Big)\Big]\, . 
\eea
The speed of sound at constant $T$ and $s/n$ as well 
as the thermal expansion coefficient are 
\bea
c_T^2 &=& \;\frac{T}{m}\, \Big[ 1 - 2zb_2(T)\Big]\, , \\
c_s^2 &=&  \frac{2T}{m}   \Big[ 1 - z
           \Big( b_2(T)+ T b_2^\prime(T)
                        + \frac{1}{2}T^2 b_2^{\prime\prime}(T)
                                            \Big)\Big]\, ,\\
\alpha &=& \;\frac{1}{T}\;  \Big[ 1 + z
           \Big( 2b_2(T) - T b_2^\prime(T)\Big)\Big]\, . 
\eea
Using these result we can compute the scale breaking parameter
$\gamma_2$
\be 
\gamma_2 = 2P- \rho c_s^2 = \frac{mT^2z^2}{\pi}
  \left( 2Tb_2^\prime (T) + T^2 b_2^{\prime\prime}(T)\right)\, . 
\ee

\section{Asymptotic expansion of $\nu(x)$}
\label{app_nu}

The asymptotic expansion of $\nu(x)$ can be found by writing
\be 
\nu(x) = \int_0^\infty dt\, \frac{\exp(-t\log(1/x))}{\Gamma(t+1)}\, , 
\ee
and Taylor expanding the gamma function, $\Gamma(t+1)^{-1}=
1+\gamma_E t + (\frac{\gamma_E^2}{2}-\frac{\pi^2}{12})t^2+
O(t^3)$. We find
\be 
\nu(x) = \frac{1}{\log(1/x)} 
       + \frac{\gamma_E}{\left[\log(1/x)\right]^2}
       + \frac{ \gamma_E^2 -\frac{\pi^2}{6}}
              {\left[\log(1/x)\right]^3} +\ldots \, . 
\ee
We can now apply this results to the thermodynamic quantities
studied in the first appendix. We find, in particular,
\bea
P-{\cal E} &=& \, \;\frac{mT^2z^2}{\pi\left[\log(T/E_B)\right]^2} 
\left\{  1   + \frac{2\gamma_E}{\log(T/E_B)} 
  + \ldots \right\}\, , \\
\gamma_2\; &=& -\frac{mT^2z^2}{\pi\left[\log(T/E_B)\right]^2} 
\left\{  1  + \frac{2\gamma_E-2}{\log(T/E_B)} 
   + \ldots \right\}\, . 
\eea

\end{appendix}


\end{document}